# Depletion of nitrogen-vacancy color centers in diamond via hydrogen passivation

A. Stacey, T.J. Karle, L.P. McGuinness, B.C. Gibson, K. Ganesan, S. Tomljenovic-Hanic, A.D. Greentree, and S. Prawer

[1]School of Physics, University of Melbourne, Parkville, Melbourne VIC 3100, Australia

R.G. Beausoleil[2]

[2]Hewlett-Packard Laboratories, 1501 Page Mill Road, Palo Alto, CA 94304 USA

A. Hoffman[3]

[3]Schulich Faculty Of Chemistry, Technion. Haifa 32000, Israel

We show a marked reduction in the emission from nitrogen-vacancy (NV) color centers in single crystal diamond due to exposure of the diamond to hydrogen plasmas ranging from 700°C to 1000°C. Significant fluorescence reduction was observed beneath the exposed surface to at least 80μm depth after ~10 minutes, and did not recover after post-annealing in vacuum for seven hours at 1100°C. We attribute the fluorescence reduction to the formation of NVH centers by the plasma-induced diffusion of hydrogen. These results have important implications for the formation of nitrogen-vacancy centers for quantum applications, and inform our understanding of the conversion of nitrogen-vacancy to NVH, whilst also providing the first experimental evidence of long range hydrogen diffusion through intrinsic high-purity diamond material.

Diamond exhibits many extreme properties, making it a material of choice for a range of advanced technological applications, such as high power/frequency electronics [1], sensitive particle detectors [2] and the nascent field of quantum-based information processing devices [3]. These applications all require precise control over impurity concentrations. Such perfection is currently only achievable with the most advanced chemical vapour deposition (CVD) synthesis techniques [4,5], which require a significant over-pressure of atomic hydrogen during synthesis[6] and typically require the generation of microwave plasmas[7].

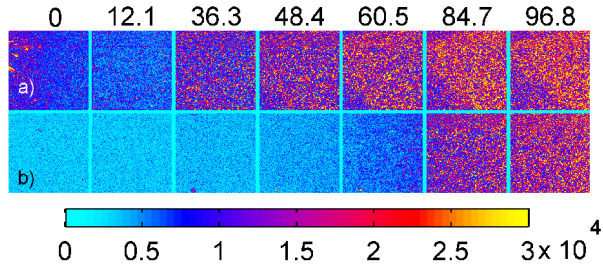

FIG 1. Fluorescent emission map in a 100μm² area for 7 different depths below the sample surface in wavelength band 650-750nm for the sample a) as received and b) post hydrogen-plasma. The number above each panel shows the depth below the sample surface in microns.

The importance of diamond for quantum information applications derives from the fact that it hosts a large variety of single photon-emitting color centers[8]. In particular, the negatively-charged nitrogen-vacancy (NV), centre is the most promising. NV is unique as it permits room temperature quantum coherence [9] and single spin single shot readout[10]. Near-term applications include the use of NV as a magnetometer[11] or as a probe of the local decoherence environment[8], and there are numerous proposals for the use of NV centers as qubits for quantum information processing[12]. Understanding the factors affecting the formation of NV is vital in ensuring access to the highest possible quality centers and maximizing yield for subsequent device integration. In this latter respect, a high yield of spectrally stable near surface NVs is crucial for coupling to monolithic and hybrid optical structures and sensitive nanoscale magnetometry. Stable, dense NV populations are now routinely observed in the bulk of high purity synthetic diamond crystals, but puzzlingly not near as-received CVD diamond surfaces. As has been previously reported[13] (and can be seen in Fig. 1a) these as-received samples often exhibit a surprising lack of NV fluorescence in the near-surface region, often extending some ~50μm into the sample.

Here we show a reduction in NV fluorescence due to plasma-induced hydrogen diffusion into diamond. This reduction informs the diffusion rate and provides an accurate lower bound on the total hydrogen trapping in the diamond. Due to the vital role of hydrogen in the growth of high-purity ("intrinsic") CVD single crystal diamond, it is likely that these results will have important implications for the growth of diamond for quantum applications.

The fluorescence emission from the as-received samples was mapped before and after plasma treatment. The single crystal diamond samples (<5ppb N, Element Six) were treated using a hydrogen plasma. Plasma conditions were similar to usual growth conditions, but without $CH_4$ in the chamber. To measure the density of emitting centres, the fluorescence was then characterised within a 100μm² area, as a function of depth, using a home built scanning confocal microscope, capable of detecting individual colour centres. Emission from the phonon side band of the NV in a 100nm wide band centered at 700nm was collected using a 0.6NA objective for an excitation laser wavelength of 532nm.

The series of confocal images shown in Fig. 1a) show the fluorescence from NV at 7 successive depths, corrected for the refractive index of diamond, for an as-received sample. Below the surface a region of about 20μm exists which exhibits reduced fluorescence when compared with the bulk. Following hydrogen plasma treatment (100Torr, 775ºC, 1.5kW, 8mins), the data in Fig.1b) show reduced fluorescence to greater depth of 80μm. The growth-relevant conditions yield a high power density plasma, stronger than typical plasmas used for hydrogen termination of the surface. It should be noted that small scratches resulting from the manufacturer's polishing process give some fluorescent signal at the surface of these samples. The hydrogen plasma is understood to slightly etch and roughen the surface, however we notice little affect upon our imaged defects in size or signal.

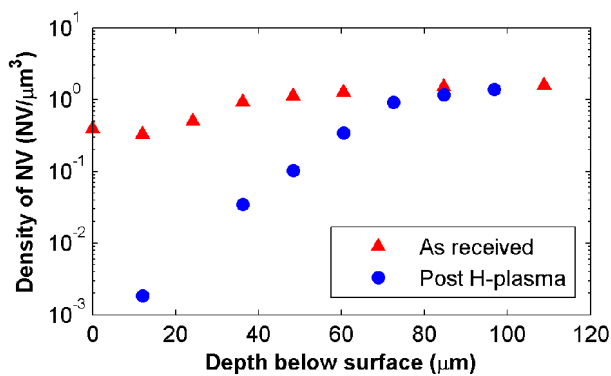

FIG 2. NV density for as-received sample measured by integrating the fluorescence over 100μm² (red triangles at 8 points below the diamond surface) and establishing the (~20kCounts/s) contribution from each emitter and estimating the confocal volume. The surface fluorescence was ignored for the as-received sample as it is dominated by surface damage and other non-NV sources. Hydrogen treated sample NV density integrated over 100μm² (blue dots) at 7 points below the diamond surface. Background corresponds to 3.5kCounts/s.

The integrated counts in these 100μm² regions are used to plot a NV density as a function of the actual depth in Fig 2, and then fitted with a complementary error function distribution, typical for gas diffusion in solid, following a 1D Fick's law. From this we were able to obtain a lowest bound estimate of the diffusion constant for hydrogen in diamond of $(0.6 \pm 0.3) \times 10^{-8}$ cm²/s. Following the plasma we obtain a mere (~3.5kCounts/s) background fluorescence signal from the sample surface. After plasma treatment a significant reduction in fluorescence is observed with the population of NV emitters recovering to the bulk crystal population at about 110μm. The observed density at this depth is consistent with the manufacturers specification of <5ppb of nitrogen, which in this case yields ~1ppt of NV.

Our results show that the hydrogen plasma treatment reduces the NV concentration by up to three orders of magnitude. Thus it can be inferred that the emitters gives an indication about the concentration of hydrogen in this sample. Other similar samples treated in this manner have also shown significant hydrogen-induced NV depletion regions >100μm below the surface (data not shown). We have been unable to observe emission from the $NV^0$ charge state in the depleted regions.

To study the influence of plasma temperature upon the diffusion a second sample was annealed in hydrogen-plasma at a pressure of 100Torr, using 1000sccm of $H_2$, and a microwave power of 1.5kW for 5 minutes at 700ºC, 800ºC and 900ºC successively. As can be seen from the data presented in Figure 3, the successive anneals steadily increases the depleted region.

A third sample treated under the same conditions as the first and then annealed at 1100°C for seven hours. The fluorescence was found not to recover, showing that the passivation of the NV is highly stable. Similarly oxygen plasma treatment was found to have no effect on the fluorescence. This is to be expected if the hydrogen is being trapped by NV to form NVH centers, which are known to be stable up towards 1600°C[14]. The expected ratio of NVH to NV in as-received high-purity diamond is around 10:1[15]. Our results suggest that hydrogen plasma significantly alters this ratio.

The reliance on hydrogen, during high-purity diamond synthesis, is known to be problematic for defect engineering purposes, due to hydrogen's tendency to decorate and passivate diamond's crystal defects. These effects have been studied extensively in the case of hybridized carbon bonding within poly-film grain boundaries[16] and for BH and PH centres in single-crystal diamond. Whilst low densities of hydrogen related defects have previously been discovered within high-purity diamonds, including various vacancy related complexes[17,18,19,20], there have been no reported direct measurements of hydrogen diffusion within these crystals. Direct measurement of hydrogen diffusion and trapping within high purity single-crystal diamond remains extremely difficult, due to the low absolute sensitivity of measurement techniques.

In non-intrinsic samples, plasma-induced hydrogen diffusion has been experimentally measured for boron rich (p-type) samples[21,22], where it is understood that diffusion of heavy hydrogen is dominated by H+ species, which are supposed to enjoy a significantly lower diffusion barrier[23], as compared to H- and H0. For intrinsic diamond however, it is generally expected that neutral hydrogen must be the dominant diffusing object[24], although there are various mechanisms and paths which hydrogen may take through the crystal lattice. Significantly, in the p-type diamond diffusion experiments, the diffusion rates and profiles were found to be dominated by trapping/detrapping phenomena, whilst implantation studies of hydrogen in single crystal diamond[25] concluded that implanted hydrogen atoms self-trap and are resistant to diffusion up to at least 1200°C.

Further information on the possibility of hydrogen diffusion within high-purity, intrinsic CVD diamond can be gleaned from the low absolute concentration of hydrogen grown into these samples, precluding the facile production of H2* under a wide range of CVD growth conditions. H-related point defects do however appear to be relatively abundant within these crystals, including the above-mentioned NVH[18] and SiVH[21] centres, with the absolute density of hydrogen within these samples appearing to scale well with the density of related H-free defects (eg. NV and SiV).

Despite these findings, there is very little other experimental information on the diffusion of hydrogen within intrinsic high-purity diamond crystals, and it is not known for example whether NVH is formed at the growing crystal surface, or the result of hydrogen combining with grown NV in the sub-surface region. It is known that electrical sample biasing during plasma treatment may enhance or suppress hydrogen diffusion[26], further highlighting the probable effects associated with changes to the local charge environment and the charge state of the hydrogen atom itself during diffusion[23]. Furthermore it is likely that Coulomb interactions between any diffusing hydrogen and its potential crystal traps play a major role. Whilst the existence of majority carriers can play a dominant role in these charge-related interactions[27], these considerations may be significantly more complicated in intrinsic samples with low-densities of defects[28].

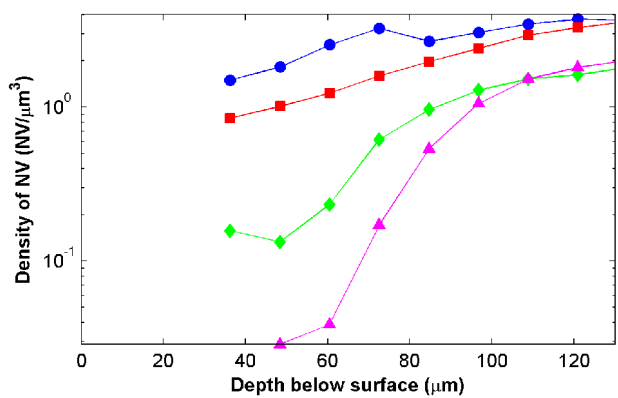

FIG 3. NV density measured by integrating over a 100μm² area for a second sample as-received (circles), and then hydrogen plasma treated at progressively higher temperatures of 700°C (squares), 800°C (diamonds) and 900°C (triangles). The surface of this sample was particularly badly scratched and thus exhibited strong non-NV fluorescence, (data not shown) .

NV is clearly and efficiently depleted within the bulk of diamond single crystals, when exposed to a high power density microwave plasma. We attribute this fluorescence depletion is due to the conversion of NV to NVH, as H+ ions are attracted to NV defects which are known to spend most of their time in the negative charge state (via coulomb interactions). It is possible that H+ represents a small proportion of the total H population within the investigated systems. The investigated temperatures of >700ºC are probably sufficient to support the diffusion of H- and H0 regardless (given maximum barrier energies of approx. 2.5eV), although they are less likely to be attracted to NV-. Due to the extreme localized Fermi-level arguments for wide band-gap insulators, it is possible that hydrogen diffuses until it is recharged into H- or H0, and it then just requires thermal excitation of charges to put it back into H+ so that it can move further into the lattice. These diffuse and stop processes can probably proceed quite rapidly, allowing for the effective high-speed diffusion of a significant flux of hydrogen within the crystal.

We demonstrate a highly sensitive measurement of the effects of hydrogen diffusing into insulating high purity (<5ppb) single crystal diamond. We spatially map the as-yet irreversible turn off of fluorescent emission from the native NV population as hydrogen diffuses through the crystal. Neither a high temperature anneal nor oxygen plasma processes were seen to reverse this process and re-establish the fluorescent population near the sample surface. Further understanding of H-diffusion and NVH formation using similar experiments are likely to play an important role in future control of NVH/NV ratios, whilst hydrogen diffusion is suggested as a potential deliberate control method for the reduction of NV densities within various applications, including the improvement of optical properties (and other defect passivations) and effective "purification" of NV centres in future QIP devices. The quenching of NV emission following H diffusion is shown to be an extremely sensitive method of probing this diffusion in an H density regime in which other hydrogen detection schemes do not typically work. Our work also demonstrates that NV concentration itself is not a robust indicator of nitrogen concentration, as NV can be turned off in the presence of hydrogen.

We acknowledge useful discussions with Daniel Twitchen and Matthew Markham at Element Six, John Goss, and Carolina Tallón for assistance with annealing. We acknowledge the Australian Research Council for financial support (project nos. DP1096288, LP100100524, DP0880466)